\documentclass[usenatbib,useAMS,hyperref]{mn2e}

\usepackage[latin1]{inputenc}
\usepackage[T1]{fontenc}
\usepackage[francais]{babel}
\usepackage{psfig,epsfig,supertabular,wrapfig,placeins}
\usepackage{graphicx,amsmath,amssymb,subfigure,multirow,units}
\usepackage{marvosym}
\usepackage{natbib}
\usepackage{euscript}
\usepackage{setspace}
\usepackage{fancyhdr}
\usepackage[usenames,dvipsnames]{color}
\usepackage{afterpage,url}
\usepackage{times}
\usepackage{float}
\usepackage{hyperref}

\hypersetup{pdftitle={On the origin of variable structures in the winds of hot luminous stars},pdfauthor={Yannick Michaux},pdfsubject={Article-MNRAS}}

\title[Variable structures in WR star winds]
      {On the origin of variable structures in the winds of hot luminous stars}
\author[Michaux et al.]
       {\parbox[t]{\textwidth}{
        Yannick J. L. Michaux$^{1, 2}$\thanks{E-mail: michaux@astro.umontreal.ca}, 
        Anthony F. J. Moffat$^{1}$,
        André-Nicolas Chené$^{3, 4, 5}$,\\
        Nicole St-Louis$^{1}$\\}
        \vspace*{3pt} \\
       $^1$Département de physique, Université de Montréal C.P. 6128, Succ. Centre-Ville, Montréal, QC, H3C 3J7 and\\
       Centre de Recherche en Astrophysique du Québec, QC, Canada\\
		   $^2$École Normale Supérieure, Lyon, CRAL, UMR CNRS 5574, Université de Lyon, France\\
       $^3$Departamento de Fisica y Astronomia, Universidad de Valparaiso, Av. Gran Breta$\tilde{n}$a 1111, Playa Ancha, Casilla 5030, Chile\\
			 $^4$Departamento de Astronomia, Universidad de Concepcion, Casilla 160-C, Chile\\
			 $^5$Gemini Observatory, Northern Operations Center, 670 North A'ohoku Place, Hilo, HI 96720, USA\\}

\begin{document}

\date{\today}

\pubyear{2012}

\maketitle


\begin{abstract}
Examination of the temporal variability properties of several strong optical recombination lines in a large sample of Galactic Wolf-Rayet stars reveals possible trends, especially in the more homogeneous WC than the diverse WN subtypes, of increasing wind variability with cooler subtypes. This could imply that a serious contender for the driver of the variations is \emph{stochastic, magnetic sub-surface} convection associated with the 170 kK partial-ionization zone of iron, which should occupy a deeper and larger zone of greater mass in cooler WR subtypes. This empirical evidence suggests that the heretofore proposed ubiquitous driver of wind variability, radiative instabilities, may not be the only mechanism playing a role in the stochastic multi small-scaled structures seen in the winds of hot luminous stars. In addition to small-scale stochastic behaviour, subsurface convection guided by a \emph{global} magnetic field with localized emerging loops may also be at the origin of the large-scale corotating interaction regions as seen frequently in O stars and occasionally in the winds of their descendant WR stars.\end{abstract}



\section{Introduction}
\label{sec:intro}
Since the first direct observational discovery of clumping in hot stellar winds \citep{Mof88}, we have still not reached a satisfactory understanding of the true physical nature of the clumps. This despite their importance in obtaining the true estimate of the mass-loss rate deviates from the first simplest estimates based on a
smooth wind. Clearly, mass-loss rates are crucial for understanding
the complete evolutionary history of hot luminous stars \citep{Mae00}. The problem of mass-loss rates applies whether
the clumps are optically thin, in which case the smooth-wind-based
mass-loss rates are overestimated by a factor of $\approx2-5$ \citep{Mof94}, or optically thick in a porous medium, which could
lead to the opposite situation \citep{Owo08}. Whatever the case,
however, it does seem apparent that clumps are a manifestation of
turbulent structures created by some kind of driver either in the
wind itself and/or external to (i.e. at the base of) the wind. Clumps
also likely lead to shocks that emit X-rays as indicated from the
simultaneous coincidence of line variations in low- versus highionization
far-ultraviolet lines in the spectrum of a key Wolf-Rayet (WR) star \citep{Mar06}.On a larger scale but related
to clumps in a broad sense are the variable discrete absorption
components (DACs) seen in the non-saturated resonant P Cygni
absorption edges of UV spectral lines of O stars \citep[e.g.][]{Mas95, Kap99}. These are related to the corotating interaction regions (CIRs) first seen in the Sun \citep{Mul84} and find their drivers on or near the rotating stellar surface probably in the form of magnetic spots or, less likely \citep[e.g.][]{Hen12} \footnote{However, \cite{Lob08} reveal that non-radial pulsations are needed to explain the second set of very slow DACs seen in the B0.5Ib supergiant HD 64760.}, standing-wave crests from non-radial pulsations \citep{Cra96}. 

The most accepted explanation for the origin of clumps is via radiative instabilities \citep{Owo88}.However,
the recent revelation that magnetic subsurface convection could be
present in hot luminous stars (primarily in OB stars evolving near
the main sequence) is certainly also a mechanism to take into consideration \citep{Can09,Can11}. When one includes the perturbative influence of the partial ionization zone of the iron elements (FeCZ) in the outer interior of the star at a temperature of close to $T_{FeCZ} \approx 170\ kK$, this could lead to acoustic and gravity waves that propagate through the outermost radiative skin \citep{Can09,Can11} and could serve as a seed to generate stochastic structures at the
base of the wind, which finally lead to the observed wind clumps.
This early seed to form clumps at the very base of the wind was
in contrast to the random fluctuations necessary to initiate the radiative
instabilities which were once believed to start further out in
the wind \citep[typically at $r \approx 1.5R_{\star}$][]{Run02} despite the observations of clumps in the wind of $\zeta$ Pup, which begin below this radius \citep{Eve98}. However, more recent work of \cite{Bou05, Bou08} reveals that UV wind line profiles down to the sonic point can only be properly fit if clumping is included. Furthermore, the improved theoretical work of \cite{Sun13} shows that photospheric perturbations and stellar limb darkening lead to instabilities starting at $r \lessapprox 1.1 R_*$.

Is there any \emph{empirical} evidence which might allow one to determine
if the subsurface convection zone contributes significantly to
the formation of clumps? In this study, we examine in this context,
the stochastic variability in the most obvious spectral lines from the
winds of stars with the strongest known stable winds, those of the
WR stars. WR stars have the distinct advantage of exhibiting strong
emission lines even in the optical (where high-precision groundbased
instruments are also very effective) from the whole wind (as
opposed to only the column towards the star in UV P Cygni absorption
edges), broadened by the Doppler effect of the (optically
thin outer) wind's rapid expansion. In a simple way, one might expect
WR stars with very high surface temperatures approaching or
above $T_{FeCZ}$ not to have significant subsurface convective zones,
while cooler WR stars would have deeper, but not too deep, FeCZ
layers, where convective driving would be more important, as for
the HeCZ in Cepheids.

While \cite{Can09,Can11} did not consider the temperature dependence of the FeCZ in W Rand He stars, we see no reason why we cannot extend their ideas for OB stars to classical WR stars, which as massive He-burning stars tend
to have (sometimes much) higher surface temperatures than those of
OB stars. In the OB-star case of \cite{Can09} it is acoustic
and gravity waves emitted in the convective zone that travel through
the radiative layer and reach the surface, thus inducing density
and velocity fluctuations. Subsurface convection zones in WR stars \cite[which have similarities to luminous main-sequence stars;][]{Can09}), have been explored and are also expected to occur near
the iron-opacity peak.While the expectation in models of subsurface
convection is for a threshold luminosity with an associated stellar
mass, the structure of realWRstars near the stellar surface where the
star ends and the wind begins is very poorly known, although some
progress has been made \citep[e.g.][]{Heg96, Pet06, Gra12}. We therefore
prefer to explore the empirical dependence of the FeCZ in WR
stars on factors that are more observationally tractable, such as the
hydrostatic surface temperature ($T_{\star}$), for which reasonably viable values are available from spectral modelling (see below). When $T_{\star}$ approaches and surpasses $T_{FeCZ} = 170\ kK$, there will be little mass involved in the subsurface convection zone to drive acoustic and gravity waves leading to wind clumping.  Conversely, as $T_{\star}$ falls below 170 kK, the FeCZ will occupy significantly moremass below
the stellar hydrostatic surface and thus could be able to drive stronger
acoustic and gravity waves, leading to stronger wind clumping. The
same driving could also lead to enhanced large-scalewind structures
such as CIRs.

On the other hand, one would expect radiative wind instabilities
to operate under a wide range of conditions, independent of the
stellar surface temperature. This would imply that if only radiative
instabilities dominate in the clump production process, then allWR
winds should be equally clumped. Alternatively, if subsurface convection
plays a role in the clump process, then the hotter stars should
show progressively less clumping. The purpose of this paper is to
test this. To do so, the basic assumption here is that the line variability
in all the stars is from clumps. Indeed, it is essential to compare
the same phenomenon in each star. Of course, if other mechanisms
contribute to the line variability, this would introduce noise in the
results. A case in point is the presence of CIRs, which could in principle
also be driven by subsurface convection leading to magnetic
fields and bright spots at the footprint of the CIRs. However, as we
discuss later, the variability of WR winds is dominated by clump
activity in most cases.
\section{Data Source}
\label{sec:data}
Fortunately, a uniform source of data to test this already exists in the literature: the set of variable line profiles of 25 northern \citep{Sai09} and 39 southern \citep{Che11} Galactic WR stars, in search for candidate CIRs. These include samples
of WR stars of all subtypes (and thus surface temperature, which
correlates well with the wind temperature or spectral subtype) in
both theWNandWC sequences. Basically, the degree of variability
was determined from four to five repeated spectra of the same star
on several different nights. Of particular importance to our present
study is the rms variation across the spectral line relative to, and thus
independent of, the local line strength, ending where the line peters
out on either side of the line centre, when the S/N becomes too
low to be significant. This is labelled as $\sigma$ in Tables \ref{table1} \& \ref{table2}, which
present all the WR stars available along with several important
stellar and wind parameters given mainly from the most recent
and up-to-date spectral analysis of Galacitc WR stars by \cite{San12} for WC/WO stars and \cite{Ham06} for WN stars, respectively. Correlations of the
line variability with these parameters will be examined in the next
section. In some cases, only upper limits are available for ?; some
of these were obtained from \cite{Che11} and three newly estimated in this paper (WR17, 23, 53). Nevertheless, the great majority of the $\sigma$ values results from calculations made from the spectra in both papers. 

Of course having only four to five spectra per star spread over
several different nights does not allow one to easily distinguish
between small-scale stochastic and large-scale periodic wind variability.
Such a distinction would require considerably more data
for each star and goes beyond the scope of this preliminary study.
However, from previous work, we can be fairly confident that most
WR-star line profile variability (LPV) is in fact dominated by clump
action, with only occasional CIR action being detected. For e.g. the study of LPV by \cite{Lep99} is based on some 20+ spectra for each of 9 WR stars, only one of which appeared to exhibit CIRs. We therefore assume that the majority of the observed LPV in most WR stars probably arises from clumps.

We separate WR stars into their two main sequences, WN and
WC, each with fundamentally different characteristics, but we deal
first withWCstars, which form a more homogeneous sequence than
WN stars.

Following the studies led by \cite{Sai09} and \cite{Che11} we take advantage of their claim that in a given star the \emph{relative} line variability is independent of the line examined, at least to first approximation.  
This is very useful, since each of these two studies did not cover the same spectral region and we had to use different lines in each case (He\begin{footnotesize}II\end{footnotesize} $\lambda$4686 or $\lambda$5411 in WN stars and C\begin{footnotesize}III\end{footnotesize} $\lambda$5696, C\begin{footnotesize}IV\end{footnotesize} $\lambda$5016 and $\lambda$5808 or C\begin{footnotesize}III\end{footnotesize}/C\begin{footnotesize}IV\end{footnotesize} $\lambda$4650 in WC stars). If more than one line was observed, we used the strongest line, which yields the more reliable result. We note that the C\begin{footnotesize}III\end{footnotesize} $\lambda$5696 line is usually more variable than other lines in WC4-8 stars, as also found in the extensive work on the WC8 star WR135 by \cite{Lep01}. This enhanced variability of C\begin{footnotesize}III\end{footnotesize} $\lambda$5696 arises from its greater sensitivity to density fluctuations \citep{Hil89}, although this does not appear to be the case for WC9 stars, where C\begin{footnotesize}III\end{footnotesize} $\lambda$5696 has a more rounded profile than its more flat-topped profiles in WC4-8 stars, implying that this line is formed relatively far out in the wind. We consider the behaviour of C\begin{footnotesize}III\end{footnotesize} $\lambda$5696 to be an anomaly  and thus not reflective of the typical variability if the (inner) wind of WC stars. Some stars in the variability study lack theoretical analysis (WR7a, 8, 21a, 79a, 79b and 83) so they were not included or considered further here. In any case, WR21a is now well established as a massive binary \citep{Nie08}. The hybrid star WR58 is taken to be WN4.  WR24 has extremely weak lines, so was not considered. WR77 has poor rectification for C\begin{footnotesize}IV\end{footnotesize} $\lambda$5808, so we take C\begin{footnotesize}III\end{footnotesize} $\lambda$5696 $\times$ 0.7 based on six other WC4-8 stars which exhibit both lines (taking limits to be actual values).
   	\begin{table*}
   		\caption{\small{Physical parameters of the WC stars. Spectral types with the suffix "d" indicate the presence of dust. $D_{Gc}$ is the distance from the Galactocentric center. $\eta$ is the effective wind opacity defined by the ratio between the rate of momentum transfer of the wind particles to photons:$\eta := \frac{\dot{M}\upsilon_{\infty}c}{L}$. $\eta = 1$ essentially means that a photon is forward scattered only once on average in its contribution to accelerate the wind.
}}
	\label{table1}
\begin{center}
\begin{tabular}{p{0.4cm}p{1.3cm}p{1cm}p{0.6cm}p{0.6cm}p{0.6cm}p{0.6cm}p{1.2cm}p{0.6cm}p{0.6cm}p{0.4cm}p{1.2cm}p{1.6cm}}
		\hline
		\hline
		WR &Type$\,^{1}$ & $\sigma\,^{3}$ & $T_{\star}\,^{2}$ & $\upsilon_{\infty}\,^{2}$ & $R_{\star}\,^{2}$ & $D_{Gc}\,^{1}$ &$\log \dot{M}_{\star}\,^{2}$ & $\log L_{\star}\,^{2}$ & $M_{\star}\,^{2}$ & $\eta\,^{2}$ & $\log L_{\star}/M_{\star}$ & Line$\,^{3}$\\
		 & &[$10^2$ \%] & [kK] &[km/s] &[$R_{\odot}$] & [kpc]&[$M_{\odot}/yr$] & [$L_{\odot}$] & [$M_{\odot}$] & & [$L_{\odot}/M_{\odot}$]&\\
		 \hline
		 4& WC5& $<$0.012&79 &2528 &2.37&9.9 &-4.68 &5.3 &12 &12.9 &4.221& CIV 5016\\
	   5& WC6& $<$0.012&79 &2120 &2.81&9.5 &-4.65 &5.45 &14 &8.3 &4.304& CIV 5016\\
		 14& WC7+?& 0.015&71 &2194 &2.98&8.3 &-4.75 &5.3 &12 &9.7 &4.221& CIV 5808\\
		 15& WC6& $<$0.012&79 &2675 &3.16&8.1 &-4.47 &5.55 &16 &12.5 &4.346& CIII 5696\\
		 17& WC5& $<$0.010 &79 &2231 &1.99&7.9 &-4.85 &5.15 &10 &11.0 &4.150& CIV 5808\\
		 23& WC6& $<$0.015 & 79 &2342 & 2.98 &7.7& -4.57& 5.5&15& 9.9 & 4.324 & CIV 5808\\
	   33& WC5& $<$0.013&79 &3342 &2.37&8.4 &-4.56 &5.3 &12 &22.5 &4.221& CIII 5696\\
		 50& WC7+OB& 0.015&71$\,^{a}$ &3200 &- &6.5&- &- &- &- &-& CIV 5808\\
		 52& WC4&$<$0.015 &112 &2765 &0.96 &7.2&-4.71 &5.12 &9 &23.3 &4.166& CIII 5696\\
	 	 53& WC8d& $<$0.015 &50 &1800 &5.00&7.0 &-4.94 &5.15 &10 &7.1 &4.150& CIV 5808\\
		 57& WC8& 0.015&63 &1787 &3.75 &6.8&-4.84 &5.3 &12 &6.4 &4.221& CIV 5808\\
		 77& WC8+OB&0.02: &60$\,^{a}$ &2300 &- &3.4&- &- &- &- & -&CIV 5808\\
	   81& WC9&0.055 &45 &1600 &6.28&6.5 &-4.7 &5.15 &10 &11.2 &4.150& CIV 5808\\
	   88& WC9&0.065 &40 &1500 &8.89&5.7 &-4.8 &5.25 &11 &6.6 &4.209& CIV 5808\\
	   90& WC7&0.009 &71 &2053 &2.75&6.5 &-4.83 &5.23 &11 &8.8 &4.189& CIV 5808\\
	   92& WC9& 0.051&45 &1121 &6.81&4.4 &-4.8 &5.22 &11 &5.3 &4.179& CIV 5808\\
	   103& WC9d+?&0.040 &45 &1190 &6.21&5.8 &-4.83 &5.14 &10 &6.2 &4.140& CIV 5808\\
	   106& WC9d&0.065 &45 &1100 &6.28&5.7 &-4.86 &5.15 &10 & 5.3&4.150& CIII 4650\\
	   111& WC5&$<$0.011 &89 &2398 &1.99 &6.5&-4.67 &5.35 &12 &11.3 &4.271& CIII 5696\\
	   119& WC9d&0.030 &45 &1300 &6.66&5.1 &-4.75 &5.2 &10 & 7.2&4.245& CIII 4650\\
	   121& WC9d& 0.050&45 &1100 &6.66&6.5 &-4.82 &5.2 & 10&5.1 &4.195& CIII 4650\\
	   135& WC8&0.010 &63 &1343 &3.66&7.7 &-4.82 &5.28 &11 &5.2 &4.239& CIII 4650\\
	   143& WC4 & 0.007:&117$\,^{a}$ & - &- &7.8&- &- &- &- &-& CIII 4650\\
	   154& WC6&$<$0.012 &79 &2300 &2.37&9.1 &-4.72 &5.3 &12 &10.7 &4.221& CIV 5016\\
		\hline
\end{tabular}
\end{center}
	 \parbox{\hsize}{{\sc Notes:}\\
															$^{1}$ From \cite{van01}.\\
	                            $^{2}$ From \cite{San12}.\\
	                            $^{3}$ From \cite{Sai09} and \cite{Che11}, with additional editing of the $\sigma$ values.\\
	                            $^a$ Interpolated values based on spectral subclass.}
\end{table*}
\begin{table*}
	\caption{\small{Physical parameters of the WN stars. Spectral types with the suffix "ha, h or (h)" indicate the presence of hydrogen emission lines sometimes with absorption from the wind. $\eta$ \& $D_{Gc}$ is as in Table 1.}}
	\label{table2}
\begin{center}
\begin{tabular}{p{0.4cm}p{1.3cm}p{1cm}p{0.6cm}p{0.6cm}p{0.6cm}p{0.6cm}p{1.2cm}p{0.6cm}p{0.6cm}p{0.4cm}p{1.2cm}p{0.4cm}p{1.4cm}}
		\hline 
		\hline
	 	WR & Type$\,^{1}$ & $\sigma\, ^{3}$ & $T_{\star}\,^{2}$ & $\upsilon_{\infty}\,^{2}$ & $R_{\star}\,^{2}$ & $D_{Gc}\,^{1}$ & $\log \dot{M}_{\star}\,^{2}$ & $\log L_{\star}\,^{2}$ & $M_{\star}\,^{2}$ & $\eta\,^{2}$ & $\log L_{\star}/M_{\star}$& $X_{H}\,^{2}$& Line$\,^{3}$\\
		 & &[$10^2$ \%] & [kK] &[km/s] &[$R_{\odot}$] &[kpc] &[$M_{\odot}/yr$] & [$L_{\odot}$] & [$M_{\odot}$] & & [$L_{\odot}/M_{\odot}$]&[\%] & \\
		 \hline
	    1 & WN4        &0.059 &112.2 &1900 &1.33&9.1 &-4.7 &5.4 &15 &7.7 &4.224&0          &HeII 4686\\
	    2 & WN2        &$<$0.008 &141.3 &1800 &0.89&9.7 &-5.3 &5.45 &16 &1.7 &4.246&0        &HeII 4686\\
	    3 & WN3ha      &0.024 & 89.1&2700 &2.65&12.6 &-5.4 &5.6 &19 &1.2 &4.321&20       &HeII 4686\\
	   10 & WN5ha      &$<$0.015 &63.1 &1100&5.61&10.7 &-5.3 &5.65 &20 &0.6 &4.349&25      &HeII 5411\\
	   18 & WN4        &$<$0.012 &112.2 &1800 &1.49&7.8 &-4.6 &5.5 &17 &6.5 &4.270&0         &HeII 5411\\
	   28 & WN6(h)+OB? & 0.02: & 50.1 & 1200 & 8.89 & 11.3 & -5.0 & 5.65 & 20 & 1.3 & 4.349 & 20  &HeII 5411\\
	   44 & WN4+OB?    &0.055 &79.4 &1400 &3.15&10.3 &-5.0 &5.55 &18 & 1.9&4.295&0     &HeII 5411\\
	   54 & WN5        &0.038 &63.1 &1500 &5.29&6.9 &-4.8 &5.6 &19 &3.0 &4.321&0           &HeII 5411\\
	   55 & WN7        &0.06 &56.2 &1200 & 8.39&6.4&-4.4 &5.8 &25 &3.5 &4.402&0            &HeII 5411\\
	   58 & WN4/WCE   &0.055& 79 & -& -&-&-& -& -& -& -& -                             &HeII 5411\\
	   61 & WN5 &0.098 &63.1 &1400 &4.21&6.2 &-4.7 &5.4 &15 &5.9 &4.224& 0          &HeII 5411\\
	   63 & WN7+OB &0.08 &44.7 &1700 &11.2 &5.8&-4.6 &5.65 &20 &5.3 &4.349& 0          &HeII 5411\\
	   67 & WN6+OB? &0.04 &56.2 &1500 &5.29&5.9 &-4.6 &5.4 &15 &6.8 &4.224& 0           &HeII 5411\\
	   71 & WN6+OB? & 0.05&56.2 &1200 &7.06&5.6 &-4.7 &5.65 &20 &2.7 &4.349&-           &HeII 5411\\
	   75 & WN6 &0.04 &63.1 &2300 &5.94&6.1 &-4.1 &5.7 & 22&19 &4.358& 0            &HeII 5411\\
	   78 & WN7h &0.02: &50.1 &1385 &16.7 &6.1&-4.2 &6.2 &44 &2.6 &4.557& 11         &HeII 5411\\
	   82 & WN7(h) &0.04 &56.2 &1100 &14.9&3.4 &-4.0 &6.3 &51 &3.1 &4.592& 20       &HeII 5411\\
	   85 & WN6h+OB? &0.03 &50.1 &1400 &21.1&3.6 &-4.2 &6.4 & 59& 1.7&4.629& 40         &HeII 5411\\
	   87 & WN7h+OB &0.02: &44.7 &1400 &18.8&5.2 &-4.6 &6.1 & 38&1.4 &4.520& 40         &HeII 5411\\
	  100 & WN7 & 0.06&79.4 &1600 &3.97&3.4 &-4.1 &5.75 &23 &12.6 &4.389& 0         &HeII 5411\\
	  108 & WN9h &0.02: &39.8 &1170 &25.1&2.6 &-4.6 &6.15 &41 &1.0 &4.537& 27       &HeII 4686\\
	  110 & WN5-6 &0.030 &70.8 &1030 &3.15&6.8 &-4.3 &5.35 &14 &23.2 &4.204& 0     &HeII 4686\\
	  115 & WN6 &0.080 &50.1 &1280 &8.89&6.1 &-4.5 &5.65 & 20& 4.3&4.349& 0        &HeII 4686\\
	  120 & WN7 &0.060 &50.1 &1225 &8.39 &5.1&-4.4 &5.6 &19 &5.7 &4.321& 0         &HeII 4686\\
	  123 & WN8 &0.050 &44.7 &970 & 17.7&4.2&-4.0 &6.05 & 35& 5.5&4.506& 0         &HeII 4686\\
	  124 & WN8h &0.050 &44.7 &710 &16.7&6.4 &-4.1 &6.0 &33 & 3.0&4.481& 13        &HeII 4686\\
	  128 & WN4(h) &0.020 &70.8 &2050 & 3.54&8.2&-5.2 &5.45 &16 &2.2 &4.246& 16    &HeII 4686\\
	  131 & WN7h &0.021 &44.7 &1400 &22.1&11.8 &-4.4 &6.3 &51 &1.3 &4.592& 20      &HeII 4686\\
	  134 & WN6 &0.060 & 63.1&1700 &5.29 &7.7&-4.4 &5.6 &19 &7.8 &4.321& 0         &HeII 4686\\
	  136 & WN6(h) &0.017 &70.8 &1600 &3.34&7.6 &-4.5 &5.4 & 15&10.9 &4.224& 12    &HeII 4686\\
	  152 & WN3(h) &0.027 &79.4 &2000 &2.23&9.0 &-5.5 &5.25 &12 &1.7 &4.171& 13    &HeII 4686\\
	  156 & WN8h &0.050 &39.8 &660 &23.7&9.8 &-4.5 &6.1 &38 &0.9 &4.520&27         &HeII 4686 \\
	  158 & WN7h &0.035 &44.7 &900&25.1&13.5 &-4.5 &6.35 &54 & 0.7&4.618& 30       &HeII 4686\\
		\hline
\end{tabular}
\end{center}
 \parbox{\hsize}{{\sc Notes:}\\
														$^{1}$ From \cite{van01} except type for WR3 \cite{Mar04}.\\
                            $^{2}$ From \cite{Ham06}.\\
                            $^{3}$ As for footnote 3 in Table 1.}\\
\end{table*}
\section{Analysis and Results}
\label{sec:res}

While the subsurface convection theory is expected to lead to a correlation
of $\sigma$ primarily with surface temperature as discussed above,
other correlations might also operate which should be examined for
the sake of completeness. We do this first for the WC stars, whose
properties have been known for some time to be more regular and
homogeneous as one advances from hot through cool subtypes than
is the case for WN stars. This may be partly due to varying amounts
of residual hydrogen in different WN stars in contrast to WC stars,
all of which have no hydrogen.  

\subsection{WC stars}
\label{sec:wc}

\begin{figure*}
   \begin{center}	  
      \includegraphics[scale=0.8]{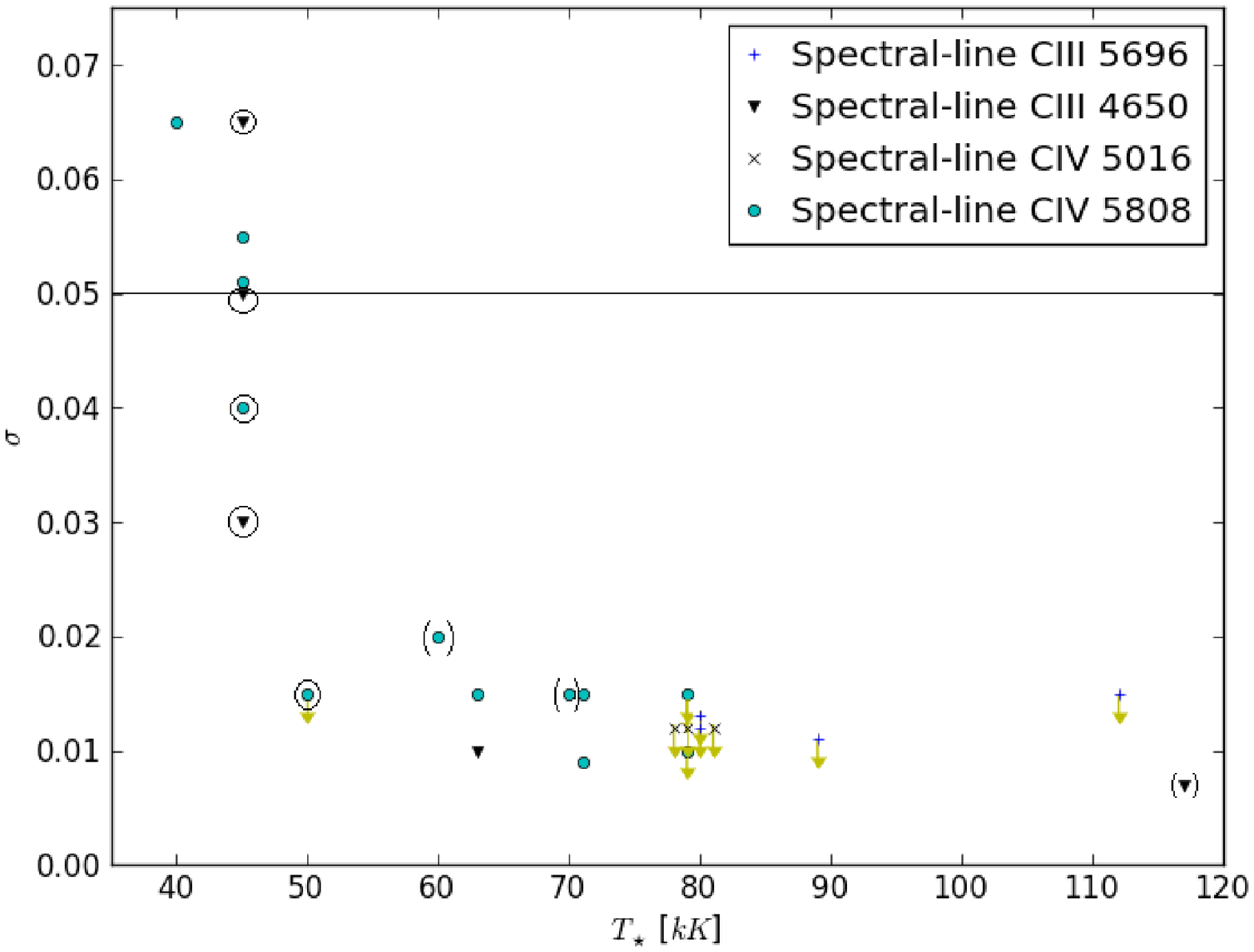}
	 \end{center}
	\caption{\small{Variability $\sigma$ for the WC stars as a function of surface temperature $T_{\star}$ for one line of each star among the spectral lines of C\protect\begin{footnotesize}III\protect\end{footnotesize} 5696 (plus signs), C\protect\begin{footnotesize}III\protect\end{footnotesize} 4650 (triangles), C\protect\begin{footnotesize}IV\protect\end{footnotesize} 5016 (crosses) or C\protect\begin{footnotesize}IV\protect\end{footnotesize} 5808 (circles). Stars that form dust are encircled. The horizontal line at $\sigma = 0.05$ separates stars selected as candidates to show CIRs (above) from the others (below), according to St-Louis et al. (2009) and Chené \& St-Louis (2011). Values for which T$_{\star}$ has been interpolated are indicated in brackets.}}
		\label{fig1} 
\end{figure*}

Fig.\ref{fig1} shows $\sigma$ vs. $T_{\star}$ taken from Table \ref{table1}, which spreads over the range 40-120 kK. Here we see a rapid increase of sigma for decreasing $T_{\star}$ below 50 kK, where all stars are of the coolest subtype, WC9. For the hotter stars, there is a gradual decrease in sigma for increasing $T_{\star}$.
We test the trend for the hotter stars by carrying out a linear correlation analysis of all non-WC9 stars. 
The dividing line set by \cite{Sai09} and \cite{Che11} between stars selected to be candidates for CIRs and the others showing variability presumably more related to the presence of clumps was arbitrarily set at $\sigma = 0.05$. It constitutes at best an
indication that the stars with a variability level above this threshold
are more likely to be found to have CIRs in their wind, which of
course, can only be proven by a detailed spectroscopic time series
with a characteristic kinematic behaviour. As a consequence, among
all WC stars only the most extreme four WC9 stars would appear
to possess CIRs. While this is possible, it seems unlikely that only
WC9 stars among WC stars should have CIRs, unless CIRs are also
driven by global magnetic subsurface convective processes which
may be stronger in WC9 stars, where the FeCZ is deeper and likely
contains more driving mass. The possible dividing line between
stars with clumps only and those with CIRs plus clumps, while highly uncertain (if such a line even exists at all), should probably be curved to follow the observed distribution more closely, i.e. dropping off at higher $T_{\star}$.

The above linear fit reveals a modest but significant correlation $r = -0.50$. A \textit{t}-test reveals that the correlation coefficient deviates significantly from zero at the 93 per cent level. Furthermore, the slope is -0.000 12(6). We have replaced the downward-pointing arrows by their most likely values of half their ordinate values. Taking them in either extreme as actual or zero does not change the results significantly.

We now look at other correlations for $\sigma$, although no plots are shown here (but see data in the Tables)\footnote{These plots for WC and WN stars are all available on request from the authors.}. 
First, $\sigma$ also varies with WC spectral subtype ($sp$) and wind terminal speed, respectively, much as seen in Fig. \ref{fig1}. 
This is certainly a consequence of the well-known relatively tight correlation for WC stars between $sp$ and $T_{\star}$ and somewhat noisier correlation between terminal speed and $T_{\star}$. 
We assume that the fundamental relation is between $\sigma$ and $T_{\star}$. The correlation between $sp$ and $T_{\star}$ shows that a hotter wind also means a hotter star.\\

Next we look at $\sigma$ vs. $R_{\star}$, mass-loss rate, 
luminosity, mass, wind opacity ($\eta = (M_{\odot}\upsilon_{\infty})/(L_{\star}/c)$) and Eddington factor $L_{\star}/M_{\star}$.
None of these shows a clear trend as seen in Fig. \ref{fig1}.\\
 
Finally we check  
for any correlation of $\sigma$ with Galactocentric distance, which is now known to correlate tightly with ambient metallicity Z \citep{Prz10} and hence initial metallicity of the star. Neglecting or not the stars with CIRs above the line at $\sigma = 0.05$ and allowing for significant dispersion in the distances,
a trend clearly emerges among the data\,\footnote{Note that the scatter in distance can be considerable, with minimum $\sigma(r)/r \approx 0.3$ for $\sigma(M_V) = 0.5$ mag for WR stars.}. The variability seems to decrease with increasing distance from the Galacic centre. 
However, any possible trend here might be related to the known fact that late-type WC stars tend to be found more frequently towards the Galactic Centre at higher Z \citep{van01}, thus making them
more variable in the context of subsurface convection (assuming that
such a zone actually does exist) affecting the wind, everything else
being equal. This is of little consequence in the context of this paper
though, since it is ultimately the surface temperature that dominates
most clearly.

It could be argued that WC9 stars seem to show peculiar behaviour
compared to other subtype WC stars, with systematically higher variability due to more frequent presence of CIRs. However,
we lack sufficient information on the presence of CIRs, which in
any case may also be more enhanced, as are clumps, in cooler stars
if the driving factor (subsurface convection) is ultimately the same
for both phenomena.

We conclude here that the only viable global correlation is between $\sigma$ and temperature, with cooler WC stars being more variable due mainly to clump (possibly with additional CIR) activity in their winds.

\subsection{WN stars}
\label{sec:wn}
\begin{figure*}
	\begin{center}
  \includegraphics[scale=0.8]{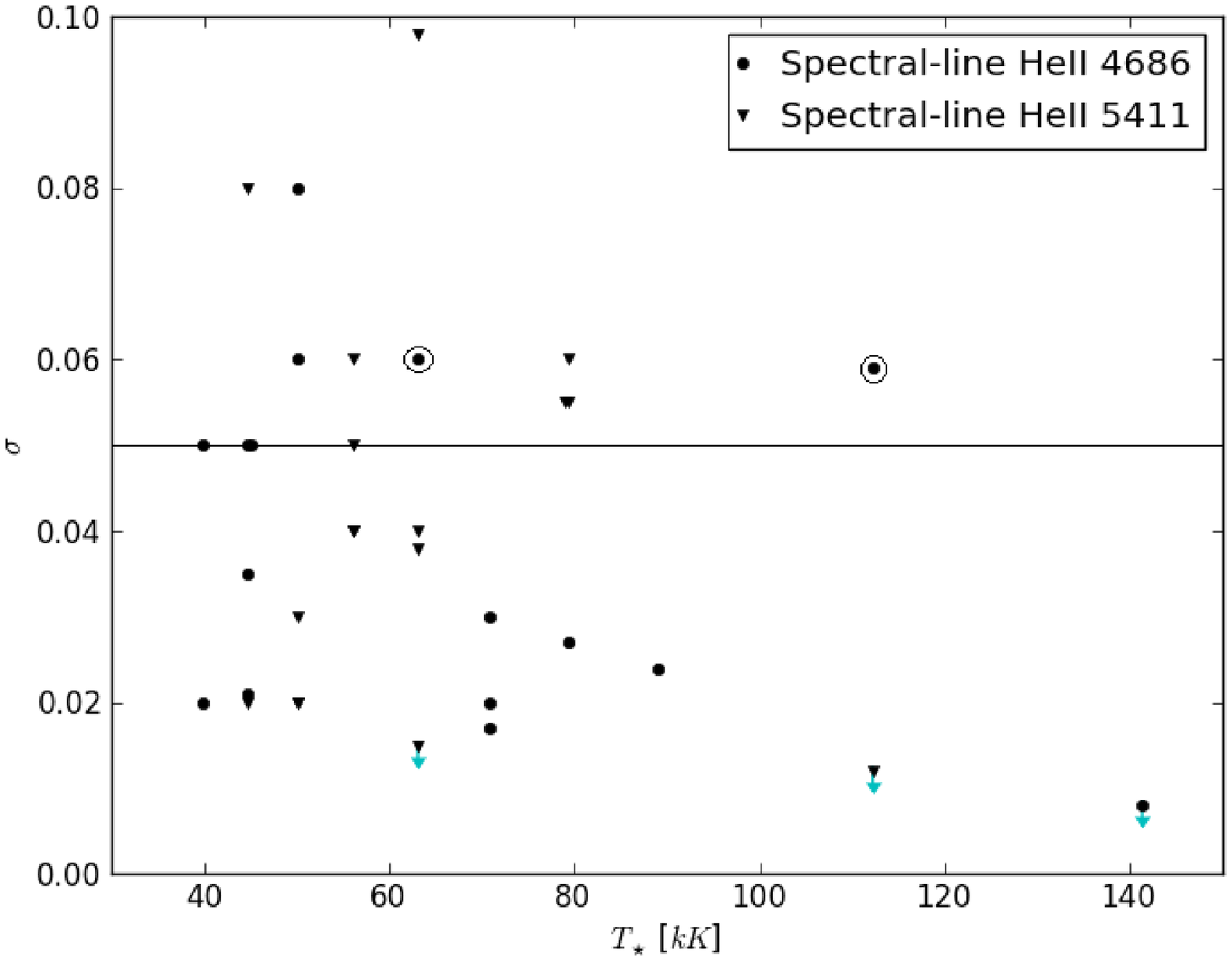}
		\end{center}
	\caption{\small{Variability $\sigma$ for the WN stars as a function of surface temperature $T_{\star}$ for one line of each star between the spectral lines of He\protect\begin{footnotesize}II\protect\end{footnotesize} 4686 (circles) or He\protect\begin{footnotesize}II\protect\end{footnotesize} 5411 (triangles). The two stars with known CIR periods, WR1 and WR134, are encircled. The horizontal line at sigma = 0.05 separates stars selected as candidates to show CIRs (above) from the others (below), according to St-Louis et al. (2009) and Chené \& St-Louis (2011).}}
		\label{fig2} 
\end{figure*}

As for the WC stars, we start in Fig. \ref{fig2} with a plot for WN stars from Table \ref{table2} of $\sigma$ vs. $T_{\star}$, which varies over the range 40-140 kK. This time, however, we do not see a clear correlation as in Fig. \ref{fig1} for the WC stars, rather a relatively large scatter in $\sigma$ at all values of $T_{\star}$. Part of this scatter could conceivably be due to the more frequent presence of CIRs in WN stars than in WC stars, thereby enhancing the values of $\sigma$ over what comes from clumps alone\,\footnote{This idea might appear to contradict the notion proposed above that clumps and CIRs could be related; however, this link need not be linear or tight.}. An indication of this comes from the two WN stars with CIRs of known period, WR134 and WR1, which are high-lighted, without however being the highest values in the figure. But we lack such information about the presence of CIRs in other stars in Fig. \ref{fig2}, making any definitive conclusion on this premature. Nevertheless, allowing for non-detections with upper limits, we do see an envelope to the scatter of points, both upper and lower, which does suggest a similar trend as in Fig.1 for the WC stars, i.e. $\sigma$ tends to be larger for the cooler subtypes. This time the division at constant $\sigma = 0.05$ between stars with possible CIRs and those possibly without, according to \cite{Sai09} and \cite{Che11} could be more realistic, although a slow drop in this division with increasing $T_{\star}$, following the envelope trend, is probably more likely.
As for the WC stars, we now look at other correlations for $\sigma$ from the WN sample. 
First, $\sigma$ varies with WN spectral subtype and wind terminal speed, respectively, 
resembling Fig.2 and is probably also a consequence of the correlations for WN stars, though noisier than for WC stars, between $sp$ and $T_{\star}$ and between terminal speed and $T_{\star}$.  
We assume that the fundamental relation is again between $\sigma$ and $T_{\star}$, even if it is quite noisy. As for WC stars the correlation between $sp$ and $T_{\star}$ for WN stars, though noisier, shows that a hotter wind also generally means a hotter star.

Next, we look at $\sigma$ vs. $R_{\star}$ and mass-loss rate. 
The former again reflects what is seen in Fig. \ref{fig2} since the radius increases with spectral subclass (although with more noise compared to WC stars). The latter only reveals a possible upper envelope, where stars with the highest $\dot{M}=dM/dt$ tend to be less variable, while the lower envelope is flat.

We now look for correlations of $\sigma$ with luminosity, mass, wind opacity ($\eta$) and Eddington factor $L_{\star}/M_{\star}$. All of these 
show no trend as seen in Fig. \ref{fig2}.
We also check for any correlation of $\sigma$ with
Galactocentric distance. Any possible trend here might be related
to the known fact that low-metallicity environments such as the
outer Galaxy and the Magellanic Clouds favour WNE among WN
stars. In this case, it is clearly impossible to determine a trend or
relation between the variability and the distance from the Galactic
Centre, which, if any exists, is masked by large dispersion.

Finally, we examine if there is a correlation between the line variability ($\sigma$) and the relative hydrogen abundance for the WN stars, $\sigma$(H). No correlation is seen, which weakens our original suspicion that variable H could be a source of the scatter in Fig. \ref{fig2}. Perhaps WN stars have more frequent CIRs of varying strength. However, $\sigma$(H) does not invalidate the hypothesis that variable H renders the WN sequence less homogeneous than the WC sequence.

We conclude here that the only viable correlation is between $\sigma$ and $T_{\star}$, with cooler WN stars being somewhat more variable due to clump and CIR activity in their winds.

\section{Discussion}
\label{sec: discuss}
According to \cite{Can09}, the strength of the convection zone in normal stars is increased for higher (ambient metallicity) $Z$, lower $T_{\star}$ and higher $L_{\star}$. Do we see this here as manifested by the degree of wind-line variability? Instead of $Z$ (difficult to estimate directly for WR stars), we looked for a proxy, Galactocentric distance $R$. As was discussed above, WC stars show a fairly clear trend, namely spectral variability decreases with increasing Galactocentric radius. This means that variability increases with increasing $Z$, thus apparently supporting Cantiello et al.'s theory. One must be prudent here though, since increasing (ambient) $Z$ also leads to cooler WC subtypes. While the impact of the convection zone and its capacity to generate stronger acoustic waves and thus stochastic structures in the wind, is metallicity dependent \citep{Can09} one might expect the variability of WN stars to be more $Z$-dependent than they are in reality, since they have not yet produced significant amounts of heavy elements beyond He, that have reached their surfaces. However, WC stars have internally produced significant C (typically $\approx 40\%$) and other heavy elements, although no Fe-peak elements, so they too should be most influenced by ambient metallicity. As for $L_{\star}$ we see no trend among WN stars and a negative trend if at all among WC stars. However, the spread in luminosity for WR stars is not large, so it is probably premature to conclude anything too significant from $L_{\star}$.

The increase in line variability (independent of the line in most cases) with decreasing model surface-temperature $T_{\star}$ is clearer for
WC stars than it is for WN stars. Some of the noise in this relation
for WN stars could be due to supplementary variability caused
by CIRs, which are strong in some WN stars and weak or undetected
in others. This seems less of a problem for WC stars, where
few if any stars deviate strongly from the mean trend. However,
the WC trend is strongly dependent on the significantly higher
variability among virtually all WC9 (and some WC8) stars. Could
this be due to some peculiarity unique to WC9 stars, which often
form dust? The answer to this is likely negative, since the
non-dust-forming WC9 stars are just as variable on average as
those that form significant dust, indicated by the subtype WC9d.
However, we cannot exclude that some other parameters might be
playing a role in rendering the WC9 stars to be more variable.
Also, additional scatter in all relationships for WN stars may be
due to their variable amounts of H [though not obvious in $\sigma$(H)], while WC stars are more homogeneous, all of them entirely lacking in H.

Standard line-driven instability should be weaker in slower, dense, optically thick WR winds \citep{Gay95}, which appears to be in contrast to what is observed.
Moreover, as recently presented by \cite{Naz13}, the low-noise X-ray light curve from the O4If star $\zeta$ Puppis can be explained by the presence of a very large number (> $10^5$) of small shocks, probably related to similar numbers of wind clumps \citep{Lep08}. This observational fact supports the existence of a convective layer as the probable origin of such a huge number of clumps \citep{Can09}, contrary to the hydrodynamical simulations based on line-driven instabilities of \cite{Fel97}, which predict many fewer X-ray shocks.

Recent work on spectral variability of WC9 (and one WC8) stars carried out by S. Desforges (in prep.), with a much larger sample of spectra than the current studies, seems to show higher values of $\sigma$ for most WC9 stars. This could be a result indeed of CIR activity, which becomes more apparent in a larger sample. However, one would also need to carry out a similar study of all WC (and WN) stars, not just WC8-9 stars before reaching any modified conclusions. We therefore take our homogeneous but limited sample as the best basis for now.

Can the trend of increased line variability with decreasing $T_{\star}$ be understood in terms of the simple analysis of WR wind clumps by \cite{Lep99}? These authors derive an expression for the relative global line-emission profile variability $\sigma_{S}/S \propto (R_w/N_e)^{1/2}$, with wind resolving power $R_w = v_e/\bar{\sigma}_{\xi}$ ($v_e =$ mean wind expansion velocity in the line-emitting region, $\propto v_{\infty}$ to good approximation, and $\bar{\sigma}_{\xi} =$ mean line-of-sight velocity dispersion) and $N_e =$ total number of elementary emitting clumps (>> $10^4$). Assuming $\bar{\sigma}_{\xi} =$ const. ($\sim100\ km s^{-1}$) for most stars, we find $\sigma_S/S \propto (v_{\infty} /N_e)^{1/2}$. If $N_e \approx$ const. \citep{Lep99}, this implies that stars with slower winds should be less variable, contrary to our finding in this work. Perhaps $N_e$ is actually smaller in cooler-wind stars? This seems unlikely, though, given the larger radii of cooler WR stars, with much larger wind volume for more clumps to develop (higher $N_e$). Although very uncertain, this scenario does not appear to be relevant, i.e. one needs something else, like temperature-sensitive subsurface convection, to be operating.

A very recent study of convective envelopes in He-rich WR stars \citep{Gra13} suggests that perturbing velocity fields at the hydrostatic surface become important in the hottest (most luminous and massive) helium stars. This is however opposite to what we observe (i.e. greater variability for cooler WR stars), possibly due to simplifying assumptions of constant (LMC) metallicity in the Grassitelli et al. study, whereas the WR stars being compared in this paper have varying Z (and thus also $T_{\star}$).

\section{Conclusions}
\label{sec:concl}

While this study of wind variability and its relation to wind structures is limited entirely to WR stars, where their strong winds make detection easier, we believe that our results may be relevant for all hot, luminous stars, where a subsurface Fe-based convection zone is now believed to be located. However, He-burning WR stars can reach much higher surface temperatures ($T_{\star}$ up to $\approx 200\ kK$) than normal H-burning stars ($T_{eff}$ up to $\approx 50\ kK$), so the effects of subsurface convection near the FeCZ at 170 kK will be better sampled. We note that O stars show no obvious trend of wind filling factor $f$ with surface temperature \citep[e.g.][]{Bou12}, although the range in $T_{eff}$ is small (c. 30-45 kK).

CIRs could also be a result of localized emerging loops in a global
magnetic field driven by subsurface convection. More work needs
to be done to determine the definitive presence of CIRs inWRstars,
by measuring their rotation periodicities and other parameters, such
as magnetism, pulsations, etc.

With the aim of supporting the work presented in this article,
it would be useful to examine WR internal structure models, in
particular with respect to density and temperature profiles in the
outer envelope where subsurface convection is expected to occur.
This would allow one to have a better idea of the importance of
the FeCZ, in terms of mass in particular, in its capacity to generate
clumps. A literature search shows that such model information is
currently not readily available.

We note that values of $T_{\star}$ are model dependent and could all be systematically in error. In fact, they are likely to be underestimates, given the theoretically higher values of surface temperatures found from internal models of He-rich massive stars \citep{Lan95}. This might explain why in this work we see the line variability starting to rise dramatically at relatively low $T_{\star}$.

In order to test the claim in this paper that subsurface convection
may be playing a significant role in hot-starwind clumps (and CIRs),
we need to enlarge the sample ofWR stars (and increase the number
of spectra per star) in search of line variability, particularly in more
of the hotter subtypes, and especially among the very hottest WR
subtypes, e.g. WO2 (WR102, WR142) with $T_{\star} = 200\ kK$. Such an effort is in progress.

\section*{Acknowledgments}

AFJM and NSL are grateful for financial support from NSERC (Canada) and FQRNT (Québec). ANC gratefully acknowledges support from the Chilean Centro de
Astrof\'isica FONDAP No. 15010003 and the BASAL Chilean Centro de Excelencia en Astrof\'isica y Tecnolog\'ias Afines (CATA) PFB-06. ANC also
received support from the Comite Mixto ESO-Gobierno de Chile and GEMINI-CONICYT No. 32110005. We are grateful to the anonymous referee for useful suggestions.

\end{document}